\documentclass[twocolumn,showpacs,preprintnumbers,amssymb,nofootinbib]{revtex4}

\usepackage{graphicx}
\usepackage{bm} 
\def\be{\begin{equation}}
\def\ee{\end{equation}}
\def\beq{\begin{eqnarray}}
\def\eeq{\end{eqnarray}}

\def\bes{\begin{eqnarray}}
\def\ees{\end{eqnarray}}
\def\half{{\textstyle{1\over2}}}
\begin{document}

\title{Scalar perturbations of higher dimensional rotating and ultra-spinning black holes}

\author{Vitor Cardoso}
\email{vcardoso@teor.fis.uc.pt} \affiliation{Centro de F\'{\i}sica
Computacional, Universidade de Coimbra, P-3004-516 Coimbra,
Portugal}

\author{George Siopsis}
\email{siopsis@tennessee.edu} \affiliation{Department of Physics and
Astronomy, The University of Tennessee, Knoxville, TN 37996-1200, USA}

\author{Shijun Yoshida}
\email{shijun@waseda.jp} \affiliation{Science and Engineering,
Waseda University, Okubo, Shinjuku, Tokyo 169-8555, Japan}

\date{\today}

\begin{abstract}
We investigate the stability of higher dimensional rotating black
holes against scalar perturbations. In particular, we make a
thorough numerical and analytical analysis of six-dimensional black
holes, not only in the low rotation regime but in the high rotation
regime as well. Our results suggest that higher dimensional Kerr
black holes are stable against scalar perturbations, even in the
ultra-spinning regime.

\end{abstract}


\maketitle
\newpage
\section{Introduction}
Exact solutions to Einstein equations are extremely useful,
specially if they describe simple yet physically attainable systems.
Indeed, take for example the famous Schwarzschild metric: with this
exact solution at hand, describing the geometry outside a
spherically symmetric distribution of matter, one was able to
compute the deflection of light as it passes near the Sun (and to
match the theoretical prediction against the observational data),
thereby giving a strong support to Einstein's theory. We now know
that the outside geometry of many astrophysical objects is well
described by the Schwarzschild metric, and we can start studying
them by investigating the properties of this metric.  One of the
most important things that one should study first is the classical
stability of a given solution.  In fact, if a solution is not
stable, then it will most certainly not be found in nature, unless
the instability timescale is much larger than the age of our
universe. What does one mean by stability? In this classical
context, stability means that a given initially bounded perturbation
of the spacetime remains bounded for all times.  For example, the
Schwarzschild spacetime is stable against all kinds of
perturbations, massive or massless \cite{regge}.  Thus, the
Schwarzschild geometry is indeed appropriate to study astrophysical
objects.  On the other hand the Kerr spacetime, describing a
rotating black hole, is stable against massless field perturbations
but not against massive bosonic fields \cite{detweiler} (although
the instability timescale is typically much larger than the age of
the Universe, thus presenting no real danger). In four dimensional
(asymptotically flat) spacetime, the Kerr-Newman family is the most
general black hole solution to Einstein equations, and, if we
exclude massive bosons, they are all stable.  However, unstable
solutions seem to be more common than previously thought. Take for
example anti-de Sitter (AdS) spacetimes. This has become a very
popular background spacetime since it was conjectured
\cite{maldacena} that there is a duality between the gravitational
degrees of freedom in the bulk of AdS space and a Conformal Field
Theory formulated on the boundary of that space; this is the AdS/CFT
correspondence conjecture.  A black hole in this spacetime
corresponds to a thermal state on the CFT. Are AdS black holes
stable? Not all, as shown recently by Cardoso and Dias
\cite{cardoso1}, who have proved that small Kerr-AdS black holes are
classically unstable. This instability is due to a ``black hole
bomb'' effect \cite{cardoso2}, whereby waves are successively
amplified near the black hole event horizon and reflected at the
boundary of the AdS spacetime (the special thing about AdS spacetime
is that its boundaries, spatial infinity, works as a wall).  If we
now consider higher dimensional spacetimes, which are of interest to
string theory and/or extra dimensional scenarios, instabilities seem
to be much more common. For instance, even though higher dimensional
Schwarzschild black holes \cite{tangherlini, myersperry} are stable
\cite{kodama}, their rotating counterparts seem not to be, at least
for large rotation. Indeed, it was proved by Gregory and Laflamme
\cite{gl} that black branes are classically unstable against a
sector of gravitational perturbations (the tensorial sector), and
this result was used recently by Emparan and Myers
\cite{myersemparan} to argue that ultra-spinning higher dimensional
black holes should be similarly unstable (recall \cite{myersperry}
that for spacetime dimensions $D$ greater than 5, $D>5$, there is no
limit on the rotation parameter). Recently Cardoso and Lemos
\cite{cardoso6} have uncovered a new universal instability for
rotating black branes and strings, which holds for any massless
field perturbation. The gist of their argument is that transverse
dimensions in a black brane geometry act as an effective mass for
the fields, which simulates a mirror enclosing a rotating black
hole, thereby creating a black hole bomb \cite{cardoso2,cardoso6}.

For other types of instabilities see for example
\cite{gibbons,marolf}.

Here, we shall investigate the stability against scalar
perturbations of ultra-spinning black holes. In four and five
dimensions, there is an upper bound for the rotation parameter $a$
of a Kerr black hole \cite{myersperry}, and when the black hole
saturates that bound we say it is an extremal black hole.  Now, it
is known \cite{leaver,cardoso3,ida} that the characteristic
frequencies (quasinormal frequencies, or QN frequencies) of four or
five dimensional Kerr black holes always have a negative imaginary
part (the field is decomposed according to $\Psi \sim e^{-i\omega
t}\Phi(r, {\rm angles})$) so the Kerr spacetime is stable. However,
as the black hole approaches extremality, the imaginary part of the
QN frequency tends to zero, thus raising the possibility that if
there was no upper bound on $a$ the QN frequencies could have a
positive imaginary part, or in other words, the spacetime could be
unstable. This will be our main motivation for this study. We are
not studying tensorial perturbations, so we shall not be dealing
with Gregory-Laflamme type of instabilities.  Instead, we are more
interested in finding out what are the consequences, if any, of
having arbitrarily large angular momentum for a black hole. We shall
focus, for concreteness, on six-dimensional rotating black holes,
but we suspect that the general features born out of this study are
valid for any spacetime dimension greater than five. Previous work
on related subject includes that of Ida {\it et al} \cite{ida} and
Berti {\it et al} \cite{berti} who studied five dimensional Kerr
black holes, in large and compact extra dimensions respectively. As
we remarked, in five dimensions there is a bound on the rotation of
the black hole, and thus these works are not suitable for studying
possibly new phenomena appearing for unbound rotation parameter.
\section{Formulation of the problem and basic equations}
\label{formulation}
\subsection{The background metric}
Here we adopt the notation of Ida et al \cite{ida}, and we also
correct some typos appearing in their equations.
In four dimensions, there is only one possible rotation axis for a
cylindrically symmetric spacetime, and there is therefore only one
angular momentum parameter.  In higher dimensions there are several
choices of rotation axis and there is a multitude of angular momentum
parameters, each referring to a particular rotation axis
\cite{myersperry}.  Here we shall concentrate on the simplest case,
for which there is only one angular momentum parameter, which we shall
denote by $a$.  The metric of a $(4+n)$-dimensional Kerr black hole
with only one non-zero angular momentum parameter is given in
Boyer-Lindquist-type coordinates by \cite{myersperry}
\begin{eqnarray}
g&=&
-{\Delta-a^2\sin^2\vartheta\over\Sigma}dt^2
-{2a(r^2+a^2-\Delta)\sin^2\vartheta \over\Sigma}
dtd\varphi \nonumber\\
&&{}+{(r^2+a^2)^2-\Delta a^2 \sin^2\vartheta\over\Sigma}\sin^2\vartheta d\varphi^2
\nonumber\\
&&{}
+{\Sigma\over\Delta}dr^2
+{\Sigma}d\vartheta^2+r^2\cos^2\vartheta d\Omega_n^2,
\label{metric}
\end{eqnarray}
where
\begin{eqnarray}
\Sigma&=&r^2+a^2\cos^2\vartheta,\\
\Delta&=&r^2+a^2-\mu r^{1-n},
\end{eqnarray}
and $d\Omega_n^2$ denotes the standard metric of the unit $n$-sphere.
This metric describes a rotating black hole in asymptotically flat,
vacuum space-time with mass and angular momentum proportional to $\mu$
and $\mu a$, respectively.  Hereafter, $\mu,a>0$ are assumed.

The event horizon is located at $r=r_H$, such that
$\Delta|_{r=r_H}=0$, which is homeomorphic to $S^{2+n}$.  For $n=0$,
the standard 4-dimensional case, an event horizon exists only for
$a<\mu/2$.  When $n=1$, an event horizon exists only when
$a<\sqrt{\mu}$, and the event horizon shrinks to zero-area in the
extreme limit $a\rightarrow\sqrt{\mu}$.  On the other hand, when $n\ge
2$, which is the part of the parameter space which we shall focus on,
$\Delta=0$ has exactly one positive root for arbitrary $a>0$.  This
means there is no bound on $a$, and thus there are no extreme Kerr
black holes in higher dimensions.

\subsection{Separation of variables and boundary conditions}
Consider now the evolution of a massless scalar field $\Psi$ in the background
described by (\ref{metric}). The evolution is governed by the
curved space Klein-Gordon equation
\be
\frac{\partial}{\partial x^{\mu}}
\left(\sqrt{-g}\,g^{\mu \nu}\frac{\partial}{\partial x^{\nu}}\Psi \right)=0\,,
\label{klein}
\ee
where $g$ is the determinant of the metric. The metric appearing in
(\ref{klein}) should describe the geometry referring to both the black
hole and the scalar field, but if we consider that the amplitude of
$\Psi$ is so small that its contribution to the energy content can be
neglected, than the Kerr metric (\ref{metric}) should be a good
approximation to $g_{\mu \nu}$ in (\ref{klein}).  We shall thus work
in this perturbative approach.  It turns out that it is possible to
simplify considerably equation (\ref{klein}) if we separate the
angular variables from the radial and time variables, as is done in
four dimensions \cite{brill}. This separation was accomplished, for
higher dimensions, in \cite{frolov} for five dimensional Kerr holes
(who work with two spin parameters) and also in \cite{page} for a
general $4+n$-dimensional Kerr hole.
Since we are considering only one angular momentum parameter, the separation is
somewhat simplified, and we can follow \cite{ida}. In the end our results agree with
the results in \cite{frolov,page}, if we consider only one angular momentum
parameter in their equations.

We consider the ansatz
$\phi=e^{i\omega t-im\varphi}R(r)S(\vartheta)Y(\Omega)$, and substitute this
form in (\ref{klein}), where $Y(\Omega)$ are hyperspherical harmonics on the $n$-sphere,
with eigenvalues given by $-j(j+n-1)$ ($j=0,1,2,\cdots$).
Then we obtain the separated equations
\begin{eqnarray}
&&{1\over\sin\vartheta\cos^n\vartheta}\left({d\over d \vartheta}
\sin\vartheta\cos^n\vartheta{dS\over d \vartheta}\right)
+\left[\omega^2a^2\cos^2\vartheta
\right.
\nonumber\\
&&{}\left.
-m^2\csc^2\vartheta
-j(j+n-1)\sec^2\vartheta
+A\right]S=0,
\label{ang}
\end{eqnarray}
and
\begin{eqnarray}
&&r^{-n}{d\over d r}\left(r^n\Delta{dR\over d r}\right)
+
\left\{
{\left[\omega(r^2+a^2)-ma\right]^2\over\Delta}
\right.
\nonumber\\
&&{}\left.
-{j(j+n-1)a^2\over r^2}
-\lambda
\right\}R=0,
\label{rad}
\end{eqnarray}
where $\lambda:=A-2m\omega a+\omega^2 a^2$.

The equations (\ref{ang}) and (\ref{rad}) must be supplemented by
appropriate boundary conditions, which are given by
\begin{equation}
R\sim\left\{
\begin{array}{ll}
(r-r_H)^{i\sigma} & {\rm as}\ r\rightarrow r_H \,, \\
r^{-(n+2)/2} {\rm e}^{-i\omega r} & {\rm as}\ r\rightarrow \infty\,.
\end{array}
\right.
\label{bound2}
\end{equation}
where
\begin{equation}
\sigma:={\left[(r_H^2+a^2)\omega-ma\right]r_H \over (n-1)(r_H^2+a^2)+2r_H^2}
\,,
\end{equation}
has been determined by the asymptotic behavior of the Eq.~(\ref{rad}).
In other words, the waves must be purely ingoing at the horizon and
purely outgoing at the infinity.  For assigned values of the
rotational parameter $a$ and of the angular index $l\,,j\,,m$ there is
a discrete (and infinite) set of frequencies called quasinormal frequencies,
QN frequencies or $\omega_{QN}$,
satisfying the wave equation (\ref{rad}) with the boundary conditions
just specified by Eq. (\ref{bound2}). The QN
frequencies are in general complex numbers, the imaginary part
describing the decay or growth of the perturbation, because the time
dependence is given by $e^{-i \omega t}$. We expect the black hole
to be stable against small perturbations, and therefore $\omega_{QN}$
is expected to have a positive imaginary part, so that the
perturbation decays exponentially as time goes by (recall that the time dependence of the
wavefunction is $e^{i\omega t}$).  As usual, we will
order the QN frequencies $\omega_{QN}$ according to the absolute value
of their imaginary part: the fundamental mode (labeled by an integer
$n=0$) will have the smallest imaginary part (in modulus), and so on.
We refer the reader to \cite{kokkotas} and \cite{cardoso4} for further details
on QN frequencies.
\section{Numerical Computation}
Now that we have a well posed problem, we have to solve for the
characteristic QN frequencies. The most powerful method to date is
that of Leaver \cite{leaver}, which makes use of a continued fraction
representation, and which can determine the resonant frequency $\omega$
and the separation constant $A$ with very high accuracy.
We assume the following series expansion for $S$
\begin{eqnarray}
S&=&(\sin\vartheta)^{|m|}(\cos\vartheta)^{j}\sum_{k=0}^\infty a_k(\cos^2\vartheta)^k,
\end{eqnarray}
which automatically satisfies the regular boundary conditions at $\vartheta=0,\pi/2$
whenever converges. Substituting this into Eq.~(\ref{ang}), we obtain the
three-term recurrence relations
\begin{eqnarray}
&&\alpha_0a_1+\beta_0a_0=0\,,\nonumber\\
&&\alpha_ka_{k+1}+\beta_ka_k+\gamma_ka_{k-1}=0\,,\quad(k=1,2,\cdots)
\label{angrec}
\end{eqnarray}
where
\begin{eqnarray}
\alpha_k&=&-2(k+1)(2j+n+2k+1)\,,\nonumber\\
\beta_k&=&(j+|m|+2k)(j+n+|m|+2k+1)-A\,,\nonumber\\
\gamma_k&=&-\omega_*^2a_*^2\,. \nonumber
\end{eqnarray}
Here, we have defined the dimensionless quantity $\omega_*:=\omega r_H$ and
$a_*:=a/r_H$, since the behavior of the system depends only on $a_*$.
When $a_*=0$, the eigenvalue $A$ is explicitly determined from the requirement that
the series expansion ends within finite terms, since otherwise divergent. Thus we have
\begin{eqnarray}
A&=&(2\ell+j+|m|)(2\ell+j+|m|+n+1)+O(\omega_*a_*),\nonumber\\
&&(\ell=0,1,2,\cdots)
\end{eqnarray}
and the 0th-order eigenfunctions are given in terms of the Jacobi polynomials:
\begin{eqnarray}
P_{\ell jm}&=&(\sin\vartheta)^{|m|}(\cos\vartheta)^{j}\nonumber\\
&&\times F\left(-\ell,\ell+j+|m|+{n+1 \over 2},j+{n+1 \over 2};\cos^2\vartheta\right).\nonumber\\
\end{eqnarray}

In a similar way, for the $n>1$ case, we expand the radial function $R$
into the form
\begin{eqnarray}
R&=&e^{-i\omega r}
\left({r-r_H\over r_H}\right)^{i\sigma}
\left({r\over r_H}\right)^{-(n+2)/2-i\sigma}
\nonumber\\
&&{}\times \sum_k^\infty b_k\left({r-r_H\over r}\right)^k,
\label{radepd}
\end{eqnarray}
where $b_0$ is taken to be $b_0=1$. If $n=2$, the expansion coefficients
$b_k$ in equation (\ref{radepd}) are determined via the
seven-term recurrence relation (it's just a matter of substituting
expression (\ref{radepd}) in the wave equation (\ref{rad})), given by
\begin{eqnarray}
&&\tilde\alpha_0b_1+\tilde\beta_0b_0=0\,, \nonumber \\
&&\tilde\alpha_1b_2+\tilde\beta_1b_1+\tilde\gamma_1b_0=0\,, \nonumber \\
&&\tilde\alpha_2b_3+\tilde\beta_2b_2+\tilde\gamma_2b_1+
\tilde\delta_2b_0=0\,, \nonumber \\
&&\tilde\alpha_3b_4+\tilde\beta_3b_3+\tilde\gamma_3b_2+\tilde\delta_3b_1+
\tilde\epsilon_3b_0=0\,, \\
&&\tilde\alpha_4b_5+\tilde\beta_4b_4+\tilde\gamma_4b_3+\tilde\delta_4b_2+
\tilde\epsilon_4b_1+\tilde\zeta_4b_0=0\,, \nonumber \\
&&\tilde\alpha_kb_{k+1}+\tilde\beta_kb_k+\tilde\gamma_kb_{k-1}+
\tilde\delta_kb_{k-2}+\tilde\epsilon_kb_{k-3} \nonumber \\
&&\quad+\tilde\zeta_kb_{k-4}+\tilde\eta_kb_{k-5}=0\,,
\quad (k=5,6,\cdots) \nonumber
\label{seven}
\end{eqnarray}
where
\begin{eqnarray}
\tilde\alpha_k&=&(1+k)(1+k+2i\sigma)(3+a_*^2)^2 \,, \nonumber \\
\tilde\beta_k&=&-18-36k^2-3\lambda-27i\sigma+36\sigma^2+18\sigma\omega_*+
               2\omega_*^2 \nonumber\\
     &-&9k(3+8i\sigma+2i\omega_*)-9i\omega_*+4a_*^3m\omega_*\nonumber \\
      &-&a_*^2\,\{12+3j(j+1)+30k^2+2m^2+\lambda+21i\sigma\nonumber\\
        &&\quad-30\sigma^2+3k(7+20i\sigma+4i\omega_*)+6i\omega_*-
       12\sigma\omega_*\} \nonumber \\
      &-&a_*^4\,\{2+j(j+1)+6k^2+4i\sigma-6\sigma^2 \nonumber \\
        &&+2k\,(2+6i\sigma+i\omega_*)+i\omega_* -2\sigma\omega_*+2\omega_*^2\}
\,, \nonumber \\
\tilde\gamma_k&=&-2m\omega_*a_*^3+a_*^4\,\{7+4j(j+1)+15k^2-10i\sigma\nonumber\\
      &&\quad\quad-15\sigma^2-2i\omega_*-8\sigma\omega_*+
        2ik(5i+15\sigma+4\omega_*)\}\nonumber \\
       &+&3\{9+20k^2+\lambda-11i\sigma-20\sigma^2-3i\omega_*-
          12\sigma\omega_*\nonumber \\
      &&\quad\quad-2\omega_*^2+ik(11i+40\sigma+12\omega_*)\}\nonumber \\
       &+&a_*^2\,\{30+9j(j+1)+62k^2+m^2+2\lambda-38i\sigma\nonumber \\
     &&\quad\quad-62\sigma^2-9i\omega_*-36\sigma\omega_*-6\omega_*^2\nonumber\\
        &&{\quad\quad}+2ik\,(19i+62\sigma+18\omega_*)\} \,, \nonumber \\
\tilde\delta_k&=&-66-54k^2-\lambda+102i\sigma+54\sigma^2\nonumber\\
    &+&6k(17-18i\sigma-5i\omega_*)+30i\omega_*+30\sigma\omega_*+
     4\omega_*^2\nonumber\\
    &-&2a_*^4\,\{14+3j(j+1)+10k^2-20i\sigma-10\sigma^2\nonumber \\
     &&\quad\quad-6i\omega_*-6\sigma\omega_*-\omega_*^2+
      2ik(10i+10\sigma+3\omega_*)\}\nonumber \\
    &-&a_*^2\,\{90+10j(j+1)+68k^2+\lambda-132i\sigma\nonumber \\
     &&\quad\quad-68\sigma^2-40i\omega_*-40\sigma\omega_*-
      6\omega_*^2\nonumber \\
        &&\quad\quad+4ik(33i+34\sigma+10\omega_*)\}
\,, \nonumber \\
\tilde\epsilon_k&=&81+28k^2-91i\sigma-28\sigma^2-21i\omega_*-
      12\sigma\omega_*\nonumber \\
    &-&\omega_*^2+ik(91i+56\sigma+12\omega_*)\nonumber \\
      &+&a_*^4\,\{47+4j(j+1)+15k^2-50i\sigma-15\sigma^2\nonumber \\
    &&\quad\quad-14i\omega_*-8\sigma\omega_*-\omega_*^2+
    2ik(2i+15\sigma+4\omega_*)\}\nonumber \\
      &+&a_*^2\,\{126+5j(j+1)+42k^2-138i\sigma-42\sigma^2\nonumber \\
     &&\quad\quad-35i\omega_*-20\sigma\omega_*-2\omega_*^2\nonumber \\
         &&{\quad\quad}+2ik(69i+42\sigma+10\omega_*)\}
\,, \nonumber \\
\tilde\zeta_k&=&-(1+a_*^2)[44+8k^2-37i\sigma-8\sigma^2-
           5i\omega_*-2\sigma\omega_*\nonumber \\
  &&\quad\quad+ik(37i+16\sigma+2\omega_*)\nonumber \\
  &&\quad\quad+a_*^2\,\{34+j(j+1)+6k^2-28i\sigma-6\sigma^2\nonumber\\
  &&\quad\quad\quad-5i\omega_*-
          2\sigma\omega_*+2ik(14i+6\sigma+\omega_*)\}]
\,, \nonumber \\
\tilde\eta_k&=&(1+a^2)^2\,(-3+k+i\sigma)^2 \,. \nonumber
\end{eqnarray}
By making a Gaussian elimination four times, one can reduce the seven-term
recurrence relations (\ref{seven}) to the three-term recurrence relations,
which is given by
\begin{eqnarray}
&&\tilde\alpha'_0b_1+\tilde\beta'_0b_0=0\,, \nonumber \\
&&\tilde\alpha'_kb_{k+1}+\tilde\beta'_kb_k+\tilde\gamma'_kb_{k-1}\,,
\ k=1,2,\cdots,
\label{three}
\end{eqnarray}
For more details on how to obtain the coefficients of the
three-term recurrence relation, we refer the reader to \cite{leaver90}.
An eigenfunction satisfying the QN mode
(QNM) boundary conditions behaves at the
event horizon and infinity as Eq. (\ref{bound2}).
Therefore, one can see that the expanded wave function (\ref{radepd})
satisfies the QNM boundary conditions if the expansion in
(\ref{radepd}) converges at spatial infinity. This convergence condition
for the expansion (\ref{radepd}), namely the QNM conditions, can be
written in terms of the continued fraction as \cite{Gu67,leaver}
\begin{eqnarray}
\tilde\beta'_0-{\tilde\alpha'_0\tilde\gamma'_1\over\tilde\beta'_1-}
{\tilde\alpha'_1\tilde\gamma'_2\over\tilde\beta'_2-}
{\tilde\alpha'_2\tilde\gamma'_3\over\tilde\beta'_3-} ...\equiv
\tilde\beta'_0-\frac{\tilde\alpha'_0\tilde\gamma'_1}{\tilde\beta'_1-
\frac{\tilde\alpha'_1\tilde\gamma'_2}
{\tilde\beta'_2-\frac{\tilde\alpha'_2\tilde\gamma'_3}
{\tilde\beta'_3-...}}} =0 \,,
\label{a-eq}
\end{eqnarray}
where the first equality is a notational definition commonly used in
the literature for infinite continued fractions. Here, we shall
adopt such a convention. As for the determination of the separation
constant $A$, exactly the same technique of the continued fraction
can be applied. The continued fraction equation for the separation
constant is then given by
\begin{eqnarray}
\beta_0-{\alpha_0\gamma_1\over\beta_1-}
{\alpha_1\gamma_2\over\beta_2-}
{\alpha_2\gamma_3\over\beta_3-} ...=0 \,.
\label{b-eq}
\end{eqnarray}
In order to obtain the QNMs, one has to solve numerically the two coupled algebraic
equations (\ref{a-eq}) and (\ref{b-eq}), following for example the procedure
in \cite{n-recipes}.
\section{Numerical Results}
\label{numerics}
\begin{figure}
\centerline{\includegraphics[width=9 cm,height=9 cm] {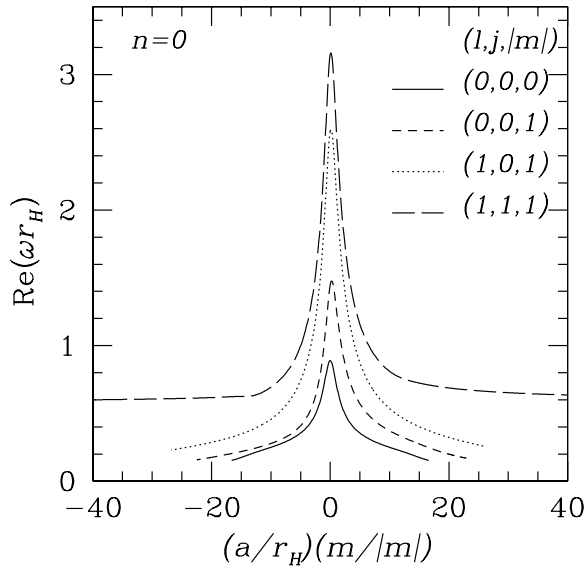}}
\caption{Real part of the fundamental QN frequency as a
function of the rotation parameter $a$ for some $l\,,j\,,m$ values.
The maximum is reached at zero rotation, and as $a$ increases the real
part of $\omega_{QN}$ decreases monotonically.
}
\label{fig:Inst1}
\end{figure}

\begin{figure}
\centerline{\includegraphics[width=9 cm,height=9 cm] {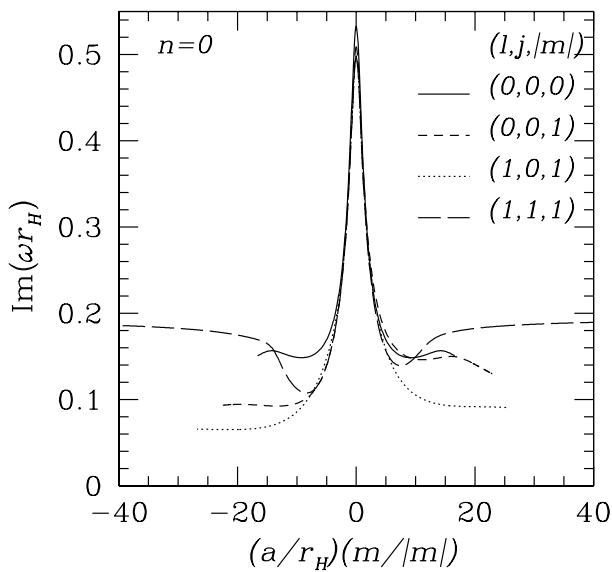}}
\caption{Imaginary part of the fundamental QN frequency as a
function of the rotation parameter $a$ for some $l\,,j\,,m$ values.
Notice that, for all values of $a$ the imaginary part is always positive,
which means that even ultra-spinning black holes are stable.}
\label{fig:Inst2}
\end{figure}

Using the method described above, we have made an extensive search
for the QN frequencies of six-dimensional rotating black holes, for
several values of the parameters $l\,,j\,,m$. The numerical results
are summarized in Table \ref{tab:1} and in Figs.
\ref{fig:Inst1}-\ref{fig:Inst2}. To check our code we have first
computed the QN frequencies of six-dimensional Schwarzschild black
holes, and compared them with analytic WKB results \cite{cardoso5}.
The results are shown in Table \ref{tab:1}, along with a computation
of the error involved in the WKB calculation.
\begin{table}
\caption{\label{tab:1} In this Table we compare our numerical
results for the QN frequencies of six-dimensional,
non-rotating black holes, with results obtained through WKB techniques, and we
also indicate the error involved using the WKB approach.
The results refer to the fundamental mode of several $l,j,m$ perturbations.}
\begin{ruledtabular}
\begin{tabular}{lllllll}  \hline
$l$&$j$&$m$ &$\omega_{QN}^{\rm Num}r_H$& $\omega_{QN}^{\rm WKB}r_H$&$\%\,{\rm Re}$&$\%\,{\rm Im}$\\ \hline
0  & 0 & 0  & 0.8894+0.5331i        & 0.7682+0.5265i         &  13.6 & 1.2      \\ \hline
0  & 0 & 1  & 1.4465+0.5093i        & 1.3846+0.4933i         &  4.3  & 3.1       \\ \hline
1  & 0 & 1  & 2.5791+0.4989i        & 2.5455+0.4942i         &  1.3  & 0.9      \\ \hline
1  & 1 & 1  & 3.1478+0.4973i        & 3.1205+0.4944i         &  0.9  & 0.6      \\ \hline
\end{tabular}
\end{ruledtabular}
\end{table}
Although for $l=j=m=0$ the error is quite large ($14\%$ for the real part) it quickly
decreases as $l\,,j\,,m$ increase. We can thus say that the code is tested (or then
that the WKB approximation yields good results...).
The full numerical results for six-dimensional rotating black holes is presented
in Figs. \ref{fig:Inst1}-\ref{fig:Inst2}.
In Fig. \ref{fig:Inst1} we present the results referring to the real part of
the fundamental QN frequency, as a function of $a$, for several $l\,,j\,,m$
values.
The real part of $\omega _{QN}$ seems to decrease monotonically as $a/r_H$
increases, and for very large $a/r_H$, it asymptotes to zero.

In Fig. \ref{fig:Inst2} we present the results referring to the
imaginary part of the fundamental QN frequency, as a function of
$a$, for several $l\,,j\,,m$ values. Although the pattern is more
complex now, one can see that up to $a/r_H =40$ the imaginary part
of $\omega _{QN}$ is still positive, and thus the modes are stable.
It proves very difficult to get numerical results for higher values
of $a/r_H$, but if the trend continues, and we have no reason to
believe otherwise, it looks like an instability will never set in.
Results for higher overtones, both for the real and imaginary part
follow a similar trend.
\section{Analytical results}
We concentrate on $n>1$ (six and higher dimensions). Let us first
re-write the radial wave equation in terms of dimensionless
variables,

\beq \label{eqwe} y^{-n} (y^n\hat\Delta R')' + \big[ \frac{(\omega_*
(y^2+a_*^2)-ma_*)^2}{\hat\Delta} - \nonumber \\
\frac{j(j+n-1)a_*^2}{y^2} - A + 2m\omega_* a_* - \omega_*^2 a_*^2
\big] R = 0 \,,\eeq where $y = r/r_H$ and $\hat\Delta = \Delta/r_H^2
= y^2+a_*^2 - (1+a_*^2) y^{1-n}$. The horizon is at $y = 1$ which is
the real root of $\hat\Delta = 0$.

In order to bring the radial wave equation into a Schr\"odinger-like
form, it is convenient to introduce \be \Psi (y) = y^{n/2}
(g(y))^{1/4} R(y)\,\,, \,g(y) = (y^2+a_*^2)^2 - a_*^2 \hat\Delta
\,.\ee In terms of $\Psi$, the wave equation~(\ref{eqwe}) reads
\be\label{eqwe2} -h(y) \left( h(y)\ \Psi' \right)' + V(y) \Psi =
(\omega_* - m\Omega_*)^2 \Psi \,,\ee where \be h(y) \equiv
\frac{\hat\Delta}{\sqrt{g(y)}}\,.\ee The potential is \beq
\frac{g(y)}{\hat\Delta}V(y) = A -2m\omega_* \Omega_* -
m^2a_*^2 \Omega_*^2+\frac{n(n+2)}{4}+\nonumber \\
\frac{[j(j+n-1)+n/2(n/2-1)]a^2}{y^2} +\frac{n^2(1+a^2)}{4y^{n+1}}
\nonumber \\+  m\Omega_* \frac{y^2-1}{\hat\Delta} [ (m\Omega_*
-2\omega_*)(y^2+a^2) + ma]  +\nonumber \\ \frac{1}{4} \left( -
\frac{5 (g')^2}{4 g^2} \hat\Delta + \frac{g''}{g} \hat\Delta +
\frac{g'}{g} \hat\Delta' \right)\,, \eeq and we have introduced the
angular velocity of the horizon, \be \Omega_H = \frac{\Omega_*}{r_H}
= \frac{a}{r_H^2 + a^2}\,. \ee The potential vanishes at the horizon
($y=1$) and approaches a constant at infinity ($V\to m\Omega_*
(m\Omega_* -2\omega_*)$ as $y\to\infty$). In general, it depends on
the frequency $\omega$. However, in the two extremal limits $a\to 0$
(Schwarzschild limit) and $a\to\infty$ (where one expects an
instability to develop), the dependence on $\omega$ drops out due to
the vanishing of the angular velocity ($\Omega\to 0$) in these
limits.

In terms of the tortoise coordinate $y_*$ defined by \be
\frac{dy}{dy_*} = h(y)\,,\ee the wave equation~(\ref{eqwe2}) may be
written as \be -\frac{d^2\Psi}{dy_*^2} + V[y(y_*)]\ \Psi =
(\omega_*- m\Omega_*)^2 \Psi \,.\ee Let us consider the two extremal
limits separately. In the limit $a\to 0$, we obtain the
Schwarzschild wave equation \be -\frac{d^2\Psi}{dy_*^2} +
V_0[y(y_*)]\ \Psi = \omega_*^2 \Psi \,,\ee where \be V_0(y) = \left(
1 - \frac{1}{y^{n+1}} \right)\ \left\{ \frac{L^2 -\frac{1}{4}}{y^2}
+\frac{(n+2)^2}{4y^{n+3}} \right\}\,, \ee  and $L= 2\ell + j +|m| +
\frac{n+1}{2}$. To estimate the eigenfrequencies, expand around the
maximum, $y_{max}$ of the potential. Setting $V_0' (y_{max}) = 0$,
we obtain \be y_{max} = \left( \frac{n+3}{2} \right)^{1/(n+1)} +
o(1/L) \,.\ee Expanding around the maximum, we may approximate the
potential by \be V_0 [ y(y_*)] \approx \alpha^2 - \beta^2
(y_*-y_*(y_{max}))^2 \,,\ee where \be \alpha^2 = V_0(y_{max}) \,,\ee
and \be \beta^2 = -\left. \half
\frac{d^2V_0}{dy_*^2}\right|_{y_*=y_*(y_{max})} = -\half
(h(y_{max}))^2V_0''(y_{max}) \,.\ee Explicitly, \beq \alpha^2 =
\frac{n+1}{n+3} \left( \frac{2}{n+3} \right)^{2/(n+1)} L^2 + o(1)
\,,\\ \beta^2 = \frac{(n+1)^3}{(n+3)^2} \left( \frac{2}{n+3}
\right)^{4/(n+1)} L^2 + o(1)\,.\eeq The wave equation becomes \be -
\Psi'' - \beta^2 x^2 \Psi = (\omega_*^2 - \alpha^2) \Psi \,,\,\, x =
y_*-y_*(y_{max})\,.\ee The solutions obeying the right boundary
conditions at $x\to\pm\infty$ are \be \Psi_N = H_N(\sqrt{i\beta} x)
e^{i\beta x^2/2} \ \ , \ \ N = 0,1,2,\dots\ee where $H_N$ are
Hermite polynomials, with corresponding eigenvalues \be \omega_*^2 =
\alpha^2 + 2i\beta (N+\half) \,.\ee Explicitly, \be \omega_* =C(n)
\left\{ L + i\sqrt{n+1} (N+\half) \right\} + o(1/L)\,, \ee with \be
C(n)=\sqrt{\frac{n+1}{n+3}}\left( \frac{2}{n+3}
\right)^{\frac{1}{n+1}}\,.\ee

This result is exactly what one gets by using a standard WKB
approach \cite{cardoso5}.

Turning to the other extreme, $a\to\infty$, we have \beq \hat\Delta
\approx a_*^2 \left( 1 - \frac{1}{y^{n-1}} \right) \,, \\
g(y) \approx \frac{a_*^4}{y^{n-1}} \,,\\ h(y) \approx y^{(n-1)/2} -
\frac{1}{y^{(n-1)/2}} \,,\eeq and the potential becomes to leading
order in $1/a$, \beq V_\infty (y) = y^{n-3} \left( 1 -
\frac{1}{y^{n-1}} \right)\times \nonumber \\ \left\{ \left(
j+\frac{n-1}{2} \right)^2 - \left( \frac{n-3}{4} \right)^2 +
\frac{(n+1)^2}{16y^{n-1}} \right\}\,, \eeq where we assumed
$A\lesssim o(a^2)$. The wave equation has a well-defined limit as
$a\to\infty$. However, the potential is positive and diverges as
$y\to\infty$ for $n>3$, so subleading terms are needed to estimate
the eigenfrequencies. For $n\le 3$ (six and seven dimensions),
$\omega$ approaches a constant value independent of $a$ which is
easily found by solving the Schr\"odinger equation. This asymptotic
value only depends on $j$.

In six dimensions ($n=2$), the potential exhibits a maximum and may
be approximated by an inverted harmonic oscillator potential, as in
the Schwarzschild limit. The frequencies can be found explicitly as
functions of $j$ taking advantage of the fact that the equation for
the maximum $V_\infty'(y_{max}) = 0$ is quadratic. In Table
\ref{tab:2} we list the QN frequencies as a function of $j$,
obtained using this analytical scheme for $a \rightarrow \infty$.
\begin{table}
\caption{\label{tab:2} In this Table we show the results of an
analytical WKB type scheme for computing the QN frequencies in the
ultra-spinning regime, $a \rightarrow \infty$. The results depend
only on $j$. This scheme shows that $\omega _{QN}$ asymptotes to a
constant value, which is consistent both qualitatively and
quantitatively with the numerical results shown in Figures
\ref{fig:Inst1} and \ref{fig:Inst2}. }
\begin{tabular}{ll}  \hline
$j$&$\hskip 2cm \omega_{QN}^{\rm Analy}r_H$ \\ \hline

0 & $\hskip 2cm 0+0.162i$\\ \hline

1  &$\hskip 2cm  0.576+0i$ \\ \hline

2  & $\hskip 2cm1.078+0i$\\
\hline
\end{tabular}
\end{table}

\section{Conclusions}
We have investigated numerically the stability of six-dimensional
rotating Kerr black holes, with one rotation parameter. Our results
suggest that this geometry is stable against scalar field
perturbations, even if the black hole is ultra-spinning. We thus
rule out the possible existence of a new kind of instability for
higher dimensional, ultra-spinning black holes. It would be
interesting to check numerically or analytically the conjecture in
\cite{myersemparan}, stating that ultra-spinning black holes should
be unstable against gravitational perturbations (more specifically,
they suggest that the Gregory-Laflamme \cite{gl} instability should
be the cause). This, for the moment, is a major challenge specially
because there is no known formalism to handle gravitational
perturbations of higher dimensional Kerr black holes. Such a
formalism could also prove useful in studying at depth the recently
discovered instability for rotating black branes and strings
\cite{cardoso6}.
\section*{Acknowledgements}
V.C. acknowledges financial support from FCT through grant
SFRH/BPD/2003. G.S. is supported in part by the US Department of
Energy under grant DE-FG05-91ER40627. S.Y. is supported by the
Grant-in-Aid for the 21st Century COE ``Holistic Research and
Education Center for Physics of Self-organization Systems'' from the
ministry of Education, Science, Sports, Technology, and Culture of
Japan.



\end{document}